\magnification=\magstephalf \baselineskip=12pt \vsize=22.0truecm
\hsize=15.5truecm
\parindent=0.8truecm
\nopagenumbers
\parskip=0.2truecm
\def\vs{\vskip 0.2in}
\def\ts{\vskip 0.05in}
\def\list{\parshape 2 0.0truecm  15.5truecm 1.15truecm 14.35truecm}

\def\o{\list$\bullet~$}
\def\n{\noindent}

\def\deg{^\circ}

\font\bigbold=cmbx12  

\centerline{{\bigbold LISA data analysis: The monochromatic binary}}

\centerline{{\bigbold detection and initial guess problems}} \vs \vs
\centerline {Ronald W. Hellings} \centerline {Department of Physics,
Montana State University, Bozeman MT 59715 }

\vs \n ABSTRACT:  We consider the detection and initial guess problems for
the LISA gravitational wave detector.  The detection problem is the
problem of how to determine if there is a signal present in instrumental
data and how to identify it.  Because of the Doppler and plane-precession
spreading of the spectral power of the LISA signal, the usual power
spectrum approach to detection will have difficulty identifying sources. A
better method must be found.  The initial guess problem involves how to
generate {\it a priori} values for the parameters of a
parameter-estimation problem that are close enough to the final values for
a linear least-squares estimator to converge to the correct result.  A
useful approach to simultaneously solving the detection and initial guess
problems for LISA is to divide the sky into many pixels and to demodulate
the Doppler spreading for each set of pixel coordinates.  The demodulated
power spectra may then be searched for spectral features.  We demonstrate
that the procedure works well as a first step in the search for
gravitational waves from monochromatic binaries.

\vfill \eject

\n {\bf I. The detection problem}

The signal coming down from a spaceborne gravitational wave detector will
contain signals from many close compact binaries, but, to look at the
signal, one would see nothing but noise.  This is because the instrument
noise, integrated over the bandwidth of the detector, far exceeds the
strength of any individual binary signal.  This situation is common in
data analysis, and the usual procedure for detecting weak coherent signals
in the presence of large incoherent noise is to generate the power
spectrum of the signal.  Since the binary star signal is monochromatic,
its power will be concentrated in a single frequency bin of the power
spectrum, while the noise is spread over all frequencies.  The signal will
then be seen as a single spike rising above the neighboring noise
spectrum.  However, the usefulness of this power spectrum method for
signal detection is limited for the spaceborne gravitational wave
detectors, due to the motion of the detector around the sun and to the
change in the orientation of the detector.  As a result of these motions,
the gravitational wave signal from a particular source will be strongly
phase modulated, spreading the total power from the source over many
frequency bins.  It is therefore quite possible that none of the frequency
components of the signal will be seen above the noise.

To illustrate, we show in Fig.\ 1 the power spectrum of the gravitational
wave signal from a binary star, where the signal frequency is 0.01 Hz and
the amplitude is $3.5\times10^{-22}$.  The signal has been sampled at 10 s
intervals for one year.  This is the power spectrum that would be seen if
the detector were at rest in the solar system.  The power spectrum as seen
in a moving detector is shown in Fig.\ 2.  To generate this figure, the
detector has been assumed to be the LISA detector [1], in which three
free-flying spacecraft move on heliocentric trajectories in such a way as
to remain at roughly constant distances from each other and to remain in a
plane inclined by $60\deg$ to the plane of the ecliptic, the plane
precessing about the ecliptic pole once per year.  The spreading of the
power over about 100 frequency bins can be clearly seen in the figure. If
a noise spectrum, characteristic of the proposed LISA instrument noise, is
added to the signal, the power spectrum in Fig.\ 3 results.  As may be
seen, the spreading of the binary signal power over many frequency bins
has produced a power spectrum that cannot now be seen above the noise.

One of the goals of this paper is to show how this ``detection problem''
may be solved.

\vs

\n {\bf II. The initial guess problem}

The ultimate goal of gravitational wave detectors is to use the
gravitational wave signals to determine the astrophysical properties of
the gravitational wave source.  Papers by Cutler and Vecchio [2,3] and by
Moore and Hellings [4] have addressed the problem of what information is
available in the signal and how well the parameters of the source may be
determined from a signal with a particular signal-to-noise ratio.  The
data analysis technique to be used to extract the information from the
signal is called linear least-squares parameter estimation.  This
technique accounts for the possible correlation of the parameters with
each other and gives the best-fit values of the parameters and their
formal standard deviation, with the assumption of a background of Gaussian
noise.  However, the ``linear'' in ``linear least-squares'' reminds us
that this procedure will only work in a totally linear problem or in the
limit that the problem may be linearized by writing it in terms of small
quantities, the differences between the final best-fit values for the
parameters and some {\it a priori} values.

The signal from a monochromatic binary is characterized by seven
independent parameters.  These are:

\o the frequency $f$ of the gravitational wave signal (a parameter that
depends on the masses of the two members of the binary and on the radius
of their mutual orbit),

\o the amplitude $h$ (which depends of the binary masses, the orbital
radius, and on the distance from the binary system to the Earth),

\o the two-parameter location of the binary in the sky (ecliptic
coordinates $\theta$ and $\phi$),

\o the two-parameter orientation of the orbit plane of the binary (taken
to be the inclination $i$ of the orbit plane normal to the line-of-sight
to the binary and the angle $\psi$ by which the major axis of the binary's
apparent ellipse in the sky is rotated relative to the ecliptic plane,
around the line-of-sight), and

\o the phase $\phi_0$ of the binary orbit at some initial time $t_0=0$.

\n The received gravitational wave signal is linear in only one of these
seven parameters, the amplitude $h$.  For all the rest, initial guesses
must be found that are sufficiently close to the correct values of the
parameters to allow the linear least-squares analysis to converge to the
proper place.  Often, in parameter estimation processes like this, the
initial guesses are known from some other aspect of the observations. For
example, in a similar problem to the one we are facing, the
times-of-arrival of pulsar pulses are used to determine the position of
the pulsar in the sky and the parameters characterizing pulsar's
rotational dynamics.  The initial guess for the position, however, is not
found from the time series itself, but from the direction in which the
antenna is pointing when the pulsar data is collected.  The spaceborne
gravitational wave detector is an omnidirectional instrument, so there
will be no such ancillary information on source direction; all information
must be extracted from the time series itself.

A critical question for LISA data analysis, therefore, is how does one
take the time series and generate initial guesses for the six non-linear
parameters.  One method would be to cross-correlate the time series with a
series of templates that represent all possible values of all possible
parameters and look for the templates that have a strong correlation with
the signal, the parameters of the template becoming the initial guess.
The spacing between templates is set by the need to get an initial guess
that is within the linear regime of the final answer. Ultimately, this is
the only method that can be employed in the absence of data from outside
the time series. However, let us consider the nature of the six non-linear
parameters to see if a more step-by-step method may be found.

The frequency $f$ of a signal buried in noise is ordinarily found by
spectral analysis.  The reason why that avenue is not available here is
evident by inspection of Figs.\ 2 and 3; the spectral power is spread out
over many frequency bins.  The contribution to this spreading by the
motion of the detector around the sun is given by the usual Doppler
formula
$$ f_r = f\left[1-{\Omega R\over c}\sin\theta \sin(\Omega t-\phi)\right]
\eqno(1)$$ where $f_r$ is the frequency observed in the detector, $f$ is
the frequency that would be observed at the solar system barycenter,
$\Omega$ is the mean motion of the detector around the sun, and $R$ is the
radius of the detector orbit. The magnitude of the Doppler spreading
depends on the gravitational wave frequency $f$ and on $\theta$.  The
phase of the Doppler depends on $\phi$.  For gravitational waves at
frequencies higher than $f=c/(2R)(\approx10^{-3}\,$Hz for LISA) this
Doppler shift dominates over the spreading produced by the precession of
the detector plane and is the major impediment to determination of the
frequency by spectral analysis. Now, if the frequency and position of a
source were to be known, the template matching could be a much simpler
procedure, with only three parameters left to determine -- $i$, $\psi$,
and $\phi_0$.  What is more, these three parameters have little effect on
the signal in the detector and are much less well-determined in the final
least-squares analysis\footnote
* {For a typical case in the binary star solutions from Moore and Hellings
[4], the uncertainties in radians in the angular parameters of a binary
star with frequency 0.01 Hz and amplitude $h=2.2\times10^{-23}$ were
$\sigma_\theta=3.68\times10^{-3}$ and $\sigma_\phi=1.18\times10^{-3}$ for
the position parameters and $\sigma_i=2.00\times10^1$ and
$\sigma_\psi=1.18\times10^2$ ({\it i.e.}, not determined) for the source
orientation parameters.} than are $\theta$ and $\phi$. Therefore the
resolution needed to find initial values for these parameters that lie
within the linear regime, where next-higher-order terms are not greater
than the final uncertainty, is very coarse, and the number of templates
needed to span the parameter space will be small.  A method to
independently find the position and frequency would thus greatly simplify
the template matching procedure.

The second goal of this paper is to show how to generate such a
step-by-step approach to the ''initial guess'' problem by approximating
the frequency and location of the source independently of the other
parameters of the system.

\vs

\n {\bf III. Demodulating the detector output}

Both problems we have discussed, the detection and initial guess problems,
may be solved by first undoing the position-dependent Doppler modulation
the signal undergoes. The Doppler modulation depends only on $\theta$ and
$\phi$ and is given simply by Eq.\ 1.  The modulation produced by the
detector-plane precession is much more complicated (see Moore and Hellings
[4]) and depends on all four angular parameters, $\theta$, $\phi$, $i$,
and $\psi$, so demodulating the Doppler only is both a simple task to
undertake and one that will undo the greatest part of the spectral
spreading at frequencies where the spreading is most pronounced.

Given a position on the sky, and given a time series from the detector, a
new time series is generated which is the time series that would be
received from that direction if the detector were at rest at the solar
system barycenter.  The plane of this fictitious detector will continue to
precess, however, at the yearly period, just like the true LISA detector.
If the time series at the detector is $y(t_n)$, we seek a time series
$z(t_n)$ that represents the signal that the barycentric detector would
have received.  The algorithm for generating $z(t_n)$ is
$$z(t_n)=y(t_n-\tau), \eqno(2)$$
where $\tau$ is the time of flight from when the wavefront passed the
detector to when it arrives at the barycenter, as seen in Fig.\ 4.  The
value of $\tau$, given $t_n$, comes from the solution of the
transcendental equation
$$\tau={\hat n\cdot\vec r(t_n-\tau)\over c},  \eqno(3)$$
where $\hat n$ is the unit vector toward the source and $\vec r(t)$ is the
vector from the barycenter to the detector at time $t$.  At time $t_n$,
the value of $\tau$ is found by solving Eq.\ 3, and the barycentric time
series is generated using Eq.\ 2.

This procedure, of course, assumes that one knows the direction $\hat n$
to the source.  When the direction is known, and the procedure applied, a
time series is generated whose Fourier spectrum is shown in Fig.\ 5. As
may be seen in comparison with Fig.\ 3, the spread spectrum that lay
beneath the noise in the uncorrected time series is now seen strongly
above the noise.  The breadth of the spectral ``line'' in Fig.\ 5 is a
result of the uncorrected modulation due to the precession of the detector
plane. Indeed, when we produce a demodulated spectrum for the OMEGA
detector [4], where the detector plane does not precess, the entire signal
power is found in a single line, as seen in Fig.\ 6.

Clearly, the remaining problem is that of determining the direction to the
source.  With no {\it a priori} knowledge, the only course of action is to
consider all directions one at a time, to demodulate for that direction,
and to examine the power spectrum of the resulting demodulated time
series.  This has the effect of solving the detection problem and the
initial guess problem simultaneously.  The demodulated signals will
collect the power into a few frequency bins where they will protrude above
the noise spectral power, the frequency at which this occurs will be the
frequency of the source, and the particular direction that produces this
visible spectral feature will be the direction to the source.

To illustrate this procedure we have simulated signals from three binary
systems, as they would be observed in the LISA detector.  We have combined
these signals together, as they would appear in the detector, and then
added random Gaussian noise at a level consistent with the LISA detector
in this frequency band.  The time series is sampled every 10 s and
accumulated for a year ($3.16\times10^6$ data points).  The sky is then
divided into pixels of one square degree, for a total of 20,582 pixels.
The pixel coordinates are used one at a time to demodulate the data
according to Eqs. 2-3, and the power spectrum is formed for each case by
fast Fourier transform of the demodulated time series. The power spectrum
is then searched for features that are a chosen factor above the average
power in the spectrum.  Since the spectral lines are still spread over
several adjoining bins by the detector plane precession, the detection
procedure takes three-point running averages of the power spectrum to
enhance such features as it searches.  The result of this procedure is
seen in Fig.\ 7.

\vs

\n {\bf IV. Discussion}

The limits of the sky plot in Fig.\ 7 are from the ecliptic plane
($\theta=90\deg$) to the ecliptic pole ($\theta=0\deg$) and over 360$\deg$
in ecliptic longitude. The negative ecliptic lattitudes ($\theta>90\deg$)
are the mirror image of the positive ones due to the symmetry of
$\sin\theta$ in Eq.\ 1.  The resolution of the plot is one square degree.
The three sources, at $\{\theta,\phi\}$ coordinates $\{60,120\}$,
$\{78,20\}$ and $\{12,280\}$ are seen clearly above the noise in the
plots.  However, the size and complicated structure of each region make it
difficult to select a value for the coordinates of the source to the same
accuracy as the resolution of the plot.  In each of the three regions, a
search for the brightest pixel for each source will give values that are
within $\pm10\deg$ of the correct values, but this is not sufficient to be
an initial guess for $\theta$ and $\phi$.  This may be seen by considering
the Doppler formula in Eq.\ 1.  The change in the Doppler shift of the
signal due to a change in $\theta$, for example, is proportional to
$$ \delta(\sin\theta)=
\cos\theta\,\delta\theta-{1\over2}\sin\theta\,\delta\theta^2 + {\rm
Higher}\, {\rm Order}\, {\rm Terms} \eqno (4)
$$
The second term on the right represents the non-linear error introduced
into the linear least-squares fit.  For the source at $\theta=60\deg$,
this non-linear term would be $1.3\times10^{-2}$ radians for
$\delta\theta=10\deg$.  This should be compared with the error in the
first term due to the final uncertainty in the linear least-squares
parameter fit,
$\cos\theta\,\sigma_\theta=(0.5)(5\times10^{-3})=2.5\times10^{-3}$ (from
the least-squares fits of Moore and Hellings [4]).  Thus, when a linear
least-squares fit tries to make a $10\deg$ change from initial guess to
final value for $\theta$, it will use only the first term in Eq.\ 4 and
will make an error five times its calculated formal uncertainty.  The
situation for $\phi$ is identical.  From this, we conclude that one cannot
go directly from the position estimates taken from the demodulated signal
to a final least-squares fit.

So what is to be done?  At this point we need to remember that our goal in
the demodulation procedure was not to generate initial guesses directly,
but to reduce the range of templates that would be needed for a
template-matching procedure.  Indeed, the complicated structure that keeps
us from determining a single accurate position via the demodulation is a
result of the precession of the detector plane that depends on $\theta$,
$\phi$, $i$, and $\psi$.  For the non-precessing OMEGA detector, the
demodulation alone is sufficient to get the initial position guess to the
required accuracy, as may be seen by examination of Fig.\ 8.  For LISA,
the result of the demodulation has only been to find an exact value for
the frequency and approximate values for $\theta$ and $\phi$.  However,
this will enable one to reduce the required template space from ($10^6$
frequencies)$\times$($2\times10^4$ positions) to less than 100 position
templates.  A five-parameter template matching procedure ($f$ is now known
and $h$ is always linear) over a restricted set of values for $\theta$ and
$\phi$ will be needed to produce a sufficient best guess to be supplied to
the final least-squares estimator.

Several other things may also be noted in Fig.\ 7.  First, a single bright
source will dominate the observed demodulated power over a large range of
values of $\theta$ and $\phi$, possibly masking weaker nearby sources.
Part of this is an artifact of the method we have used to generate Fig.\
7, the picking out of the strongest spectral feature to represent each
pixel. Perhaps the wisest use of the figure is to eliminate regions where
there is no significant demodulated power as being a region free of
sources that reach above the noise at all.  One may then concentrate on
the regions with significant power and examine spectra from many points in
the region to look for secondary sources with different frequencies.  A
second thing to be noticed Fig.\ 7 is the significant leaking of power
from sources near the ecliptic plane ($\theta=90\deg$) down towards the
ecliptic plane. This is a result of the diminished position accuracy for
the least squares fits for monochromatic binaries as one approaches the
ecliptic (see Moore and Hellings [4], Fig.\ 3).  Looking at it another
way, this is a result of the fact that the phase of the Doppler modulation
is the same for sources at the same ecliptic longitudes; only the
amplitude of the modulation is different.

Finally, let us consider some of the questions that remain to be
addressed.  First, the details of the final template matching need to be
studied.  Cornish and Larson [5] have generated an efficient code for
template matching that works in the frequency domain and that is capable
of fast correlation of signals over the entire parameter space.  They have
also generated a Doppler demodulation code[6], that similarly works in the
frequency domain, whose efficiency should be compared with the time-domain
code that we have used in this paper.  Lastly, we should point out that
among the test cases we have run are some in which there are two or more
sources nearby each other in frequency and in position.  When this occurs,
there seems to be a pulling of the best demodulated pixels away from both
source positions to some intermediate position.  This has implications not
only for the initial guess problem, but also for the problem of parameter
estimation in the presence of non-Gaussian noise.  These are issues we
intend to address in future papers.

It is a pleasure to thank Neil Cornish and Thomas Moore for helpful
discussions.  This study was accomplished with support from NASA grants
NAG5-11469 and NCC5-579.

\vfill

\eject

\n {\bf References:}

\ts

\font\bo=cmbx10

\n \hangindent=0.2in \hangafter=1 [1] K.\ Danzmann {\it et al.,} {\it LISA
Pre-Phase A Report (second edition)}, MPQ 233 (1998).

\n \hangindent=0.2in \hangafter=1 [2] C.\ Cutler and A.\ Vecchio, {\it
Proceedings of the 2$^{nd}$ LISA Symposium (AIP Conference Proceedings
456)}, 95 (1998).

\n \hangindent=0.2in \hangafter=1 [3] A.\ Vecchio and C.\ Cutler, {\it
Proceedings of the 2$^{\rm nd}$ LISA Symposium (AIP Conference Proceedings
456)}, 101 (1998).

\n \hangindent=0.2in \hangafter=1 [4] T.A.\ Moore and R.W.\ Hellings,
Phys.\ Rev.\ {\bo D65} 062001 (2002).

\n \hangindent=0.2in \hangafter=1 \n [5]  N.J.\ Cornish and S.L.\ Larson,
manuscript in preparation (2002).

\n \hangindent=0.2in \hangafter=1 \n [6]  N.J.\ Cornish and S.L.\ Larson,
manuscript in preparation (2002).

\vfill

\eject

\n CAPTIONS:

\n Figure 1.  A portion of the power spectrum of the gravitational wave
from a binary star, as seen by a stationary, non-precessing detector at
the barycenter.  The parameters characterizing the gravitational wave
source are $h=3.5\times10^{-22}$, $f=1.0\times10^{-2}\,$Hz,
$\theta=60\deg$, $\phi=120\deg$, $i=10\deg$, $\psi=84\deg$, and
$\phi_0=0$.

\n Figure 2.  The same gravitational wave as in Fig.\ 1, as seen by the
LISA detector.  The power is now spread over about one hundred frequency
bins.

\n Figure 3.  The same gravitational wave in the LISA detector as in Fig.\
2, but with noise added at a level characteristic of the LISA detector in
this frequency band.  The spread spectrum cannot now be detected above the
noise.

\n Figure 4.  The geometry of the revision of the wavefront timing to the
solar system barycenter.  In the first picture, the LISA detector receives
the signal at time $t=t_n-\tau$.  In the second picture, a time $\tau$
later, the same signal is received at the barycenter, while LISA has moved
to a new position.

\n Figure 5.  The signal from Fig.\ 3, after the revision to the solar
system barycenter has been performed for the particular direction of this
source.  The signal at 0.01 Hz is clearly visible above the noise.  The
remaining breadth of the spectral peak is due to the modulation produced
by the precessing plane of the LISA detector.

\n Figure 6.  The same as Fig. 5, but for the non-precessing OMEGA
detector.  The entire gravitational wave power is now concentrated in a
single peak.

\n Figure 7.  The all-sky map of peak spectral power in the demodulated
signal for one-degree pixels.  The signal being analyzed represents the
received LISA time series for three binary gravitational wave sources plus
random noise at the predicted LISA level. The parameters of the three
sources are: SOURCE 1 --- $h=3.3\times10^{-22}$, $f=3.0\times10^{-3}\,$Hz,
$\theta=60\deg$, $\phi=120\deg$, $i=10\deg$, $\psi=84\deg$, $\phi_0=0$;
SOURCE 2 --- $h=4.0\times10^{-22}$, $f=4.0\times10^{-3}\,$Hz,
$\theta=78\deg$, $\phi=20\deg$, $i=40\deg$, $\psi=0\deg$, $\phi_0=0$; and
SOURCE 3 --- $h=3.7\times10^{-22}$, $f=3.5\times10^{-3}\,$Hz,
$\theta=12\deg$, $\phi=280\deg$, $i=0\deg$, $\psi=84\deg$, $\phi_0=0$.

\n Figure 8.  The same type of all-sky map for the same signals as Fig. 7,
but for the non-precessing OMEGA detector.  The relative weakness of the
power near the ecliptic (the $\theta=78\deg$, $\phi=20\deg$ location) is
due to the non-uniform sensitivity of the OMEGA detector -- more sensitive
than LISA near the ecliptic pole and less sensitive near the ecliptic
plane.

\bye